\documentclass{article}

\usepackage[preprint]{neurips_2026}

\usepackage[utf8]{inputenc} % allow utf-8 input
\usepackage[T1]{fontenc}    % use 8-bit T1 fonts
\usepackage{hyperref}       % hyperlinks
\usepackage{url}            % simple URL typesetting
\usepackage{booktabs}       % professional-quality tables
\usepackage{amsfonts}       % blackboard math symbols
\usepackage{nicefrac}       % compact symbols for 1/2, etc.
\usepackage{microtype}      % microtypography
\usepackage{xcolor}         % colors
\usepackage{wrapfig} 
\usepackage{graphicx}
\usepackage{xcolor}
\usepackage{tcolorbox}
\usepackage{amsmath}
\usepackage{amssymb}
\usepackage{algorithm}
\usepackage{algorithmic}
\title{``Theater of Mind'' for LLMs: A Cognitive Architecture Based on Global Workspace Theory}

\author{%
  Wenlong~Shang\thanks{Code: \url{https://github.com/giansha/Global-Workspace-Agents}} \\
  Beijing Key Laboratory of Computational Intelligence and Intelligent System\\
  Beijing University of Technology\\
  Beijing, China \\
  \texttt{shangwl@emails.bjut.edu.cn}
}

\begin{document}
\maketitle

\begin{abstract}
Modern Large Language Models (LLMs) operate fundamentally as Bounded-Input Bounded-Output (BIBO) systems. They remain in a passive state until explicitly prompted, computing localized responses without intrinsic temporal continuity. While effective for isolated tasks, this reactive paradigm presents a critical bottleneck for engineering autonomous artificial intelligence. Current multi-agent frameworks attempt to distribute cognitive load but frequently rely on static memory pools and passive message passing, which inevitably leads to cognitive stagnation and homogeneous deadlocks during extended execution. To address this structural limitation, we propose Global Workspace Agents (GWA), a cognitive architecture inspired by Global Workspace Theory. GWA transitions multi-agent coordination from a passive data structure to an active, event-driven discrete dynamical system. By coupling a central broadcast hub with a heterogeneous swarm of functionally constrained agents, the system maintains a continuous cognitive cycle. Furthermore, we introduce an entropy-based intrinsic drive mechanism that mathematically quantifies semantic diversity, dynamically regulating generation temperature to autonomously break reasoning deadlocks. Coupled with a dual-layer memory bifurcation strategy to ensure long-term cognitive continuity, GWA provides a robust, reproducible engineering framework for sustained, self-directed LLM agency.
\end{abstract}

\section{Introduction}
\label{sec:introduction}

% \begin{wrapfigure}{r}{0.35\textwidth}
% \centering
% \includegraphics[width=0.35\textwidth]{example-image-duck}
% \caption{The classical BIBO system paradigm.}
% \label{fig:bibo}
% \end{wrapfigure}

In classical control theory, a Bounded-Input Bounded-Output (BIBO) system is defined as an entity that produces a finite, strictly bounded output in direct response to a provided input. Modern Large Language Models (LLMs) and their corresponding single-agent architectures fundamentally operate as advanced BIBO systems. They remain in a dormant state until explicitly triggered by a user prompt, at which point they compute a localized response and immediately return to passivity. While this reactive paradigm demonstrates high efficacy for executing isolated instructions, it presents a critical limitation when engineering genuinely autonomous artificial intelligence. An autonomous system must possess the capacity for self-initiated action and continuous internal deliberation, rather than merely acting as a passive function call awaiting external stimulation. This architectural limitation prompts a fundamental inquiry: how might we organize one or more of these BIBO LLM agents to transcend their individual reactivity and synthesize a continuously operating autonomous system?

% \begin{wrapfigure}{l}{0.35\textwidth}
% \centering
% \includegraphics[width=0.35\textwidth]{example-image-duck}
% \caption{Cognitive stagnation in closed loops.}
% \label{fig:stagnation}
% \end{wrapfigure}

To explore this constraint, consider a straightforward closed-loop thought experiment. Suppose we connect two identical BIBO agents in a cyclical configuration, where the output of the first agent serves directly as the input for the second, and vice versa. Intuitively, one might expect this topological loop to generate a self-sustaining cycle of continuous reasoning. However, empirical observation reveals that the system rapidly descends into a state of cognitive stagnation. Recent studies on multi-agent interactions characterize this stagnation through phenomena such as sycophancy, echo chambers, and the degeneration of thought \cite{TalkIsntCheap_FailureModes_2025, CONSENSAGENT_Sycophancy_2025}. In these homogeneous setups, agents inherently favor social conformity over critical engagement, frequently converging on repetitive, trivial agreements or mutually reinforcing hallucinated logic \cite{MultiAgent_Consensus_2023}. The root cause of this stagnation lies in their homogeneity and the absence of an intrinsic computational drive. Because neither agent possesses the architectural mandate to introduce radical novel information, evaluate global progress, or unilaterally shift the cognitive trajectory, the head-to-tail loop quickly exhausts its initial semantic momentum. Without a centralized mechanism to orchestrate diversity, the multi-agent setup inevitably collapses back into localized loops.

To contextualize this deadlock, we must critically examine the rapidly evolving paradigms in agentic reasoning. Foundational methodologies, most notably Chain-of-Thought \cite{wei2022chain_cot} and ReAct \cite{yao2022react}, have significantly advanced the reasoning capabilities of individual LLMs by interleaving internal thought generation with external task execution. To overcome the strict linearity of these approaches, subsequent frameworks such as Tree of Thoughts \cite{Yao_TreeOfThoughts_2023} and Graph of Thoughts \cite{Besta_GraphOfThoughts_2023} introduced non-linear exploration of reasoning paths. Furthermore, self-correction mechanisms endowed individual agents with iterative refinement capabilities based on environmental feedback \cite{Shinn_Reflexion_2023}. Nevertheless, these paradigms remain fundamentally reactive and computationally isolated. They rely on a single cognitive stream that is highly susceptible to context degradation over extended reasoning trajectories.

To distribute this cognitive load, researchers have actively explored multi-agent frameworks. Role-playing methodologies, such as CAMEL \cite{Li_CAMEL_2023}, alongside communicative frameworks like AutoGen \cite{wu2024autogen} and MetaGPT \cite{hong2023metagpt}, facilitate decentralized task execution through sequential message passing. Concurrently, multi-agent debate paradigms leverage distinct personas to cross-examine reasoning flaws, thereby reducing localized biases \cite{Du_MultiAgentDebate_2023}. While these systems demonstrate enhanced performance on complex tasks, they predominantly rely on static shared memory pools or peer-to-peer dialog trees. As established in our thought experiment, passive message passing between statically assigned nodes fails to maintain global semantic coherence. Without an intrinsic drive mechanism to evaluate the overarching state, these systems cannot foster true spontaneous agency and frequently require external human intervention to prevent conversational truncation.

Recent attempts to bridge this gap have drawn inspiration from cognitive science, notably the instantiation of long-term memory and social dynamics in generative environments \cite{Park_GenerativeAgents_2023}, or the conceptual mapping of classical cognitive architectures onto modern LLMs \cite{Sumers_CognitiveArchitecturesLLMs_2023}. Yet, a robust computational implementation of a consciousness-like centralized broadcasting hub remains largely unexplored. 

To overcome these structural limitations, an autonomous architecture requires two foundational elements: strict cognitive heterogeneity and an active, centralized information broadcasting mechanism. In this work, we propose Global Workspace Agents (GWA), an architecture inspired by Global Workspace Theory \cite{baars1997theatre}. Rather than relying on static memory pools, GWA implements a dynamic cognitive cycle. By coupling a central broadcast hub with a diverse set of functionally specialized agents, the system mitigates the reactive constraints inherent in previous frameworks. This approach directly addresses the homogeneous deadlock, providing a mathematically grounded and verifiable engineering framework for LLM-based systems to maintain continuous, self-directed reasoning.

\section{From Blackboard to the Global Workspace}
\label{sec:theoretical_foundations}

The evolution of multi-agent architectures relies fundamentally on the mechanisms governing inter-agent communication. Conventional frameworks frequently coordinate agents through static shared memory or blackboard architectural patterns \cite{craig1988blackboard}. In these paradigms, memory functions as a passive data structure. Agents must explicitly poll the shared state to determine subsequent actions. This passive design presents a computational bottleneck for autonomous systems: without external triggers or hardcoded sequential pipelines, the agents lack a structural mechanism to spontaneously initiate tasks or interrupt ongoing processes.

To address this limitation, we draw theoretical grounding from cognitive science, specifically the Global Workspace Theory (GWT) proposed by Baars \cite{baars2005global}. GWT offers a structural model for how parallel, decentralized processes can synthesize unified, coherent behavior.

\subsection{Demystifying the ``Theater of Mind'' Metaphor}

% \begin{figure}[htbp]
%     \centering
%     \includegraphics[width=0.6\textwidth]{example-image-duck}
%     \caption{The ``Theater of Mind'' cognitive model, illustrating the dynamic interplay between the global workspace and specialized unconscious processors.}
%     \label{fig:theater_of_mind}
% \end{figure}

A core pedagogical construct in GWT is the ``Theater of Mind'' metaphor \cite{baars1997theatre}. Rather than treating memory as a passive storage drive, Baars conceptualizes cognitive architecture as a dynamic functional state. The namesake ``Global Workspace'' is not a discrete anatomical structure, but rather an emergent broadcast hub. The model is built upon three primary interacting components:

\begin{itemize}
    \item \textbf{The Stage (Working Memory):} A centralized, limited-capacity integration domain where diverse information streams are temporarily held.
    
    \item \textbf{The Spotlight (Attention):} A rigorous selection mechanism that highlights highly specific information on the stage. The illuminated subset of this stage constitutes the actual Global Workspace. Any information entering this spotlight achieves global visibility.
    
    \item \textbf{The Audience (Specialized Processors):} A vast collection of decentralized, specialized cognitive networks operating in parallel within the unconscious periphery.
\end{itemize}

The fundamental mechanism driving this cognitive model is the global broadcast. When the spotlight illuminates specific data on the stage, elevating it into the Global Workspace, that information is instantaneously transmitted to the entire audience. The specialized processors evaluate this broadcast based on their strictly defined functional domains. If a processor identifies relevant data, it activates, processes the information, and subsequently attempts to push its own computational output back onto the stage, competing for attentional resources.

\subsection{Mapping GWT to LLM Agent Topologies}

Translating this cognitive framework into a verifiable multi-agent engineering architecture requires a precise functional mapping \cite{dehaene2011global}. In our Global Workspace Agents (GWA) architecture, we replace passive shared memory with an active, event-driven broadcasting hub.

\begin{itemize}
    \item \textbf{The Stage} is engineered exclusively as the Short-Term Working Memory (STM). It functions as a centralized state tensor that maintains the immediate dialogue context and the current global state configuration.

    \item \textbf{The Spotlight} is instantiated as the Attention Node. It bridges the immediate context on the stage with the vast repository of Long-Term Memory (the off-stage archives), actively selecting which contextual vectors are elevated to global visibility.

    \item \textbf{The Audience} is implemented as a swarm of heterogeneous LLM agents. Each agent is initialized with a distinct instructional invariant dictating a specific cognitive role, such as divergent generation, logical critique, or executive arbitration.
\end{itemize}

The critical architectural shift is the implementation of the broadcast mechanism. When a state update occurs in the Global Workspace, the new information tensor is explicitly broadcast to all connected nodes. This broadcast acts as an event-driven activation signal. Each agent continuously evaluates this global state. When an agent determines that the broadcasted context requires its specific computational expertise, it initiates internal processing and submits a structured proposal back to the workspace. This event-driven routing effectively mitigates the limitations of passive data structures, providing a functional foundation for spontaneous and continuous multi-agent interaction.

\section{The Global Workspace Agents Architecture}
\label{sec:architecture}

% \begin{figure}[htbp]
%     \centering
%     \includegraphics[width=0.6\textwidth]{example-image-duck}
%     \caption{The architecture of Global Workspace Agents.}
%     \label{fig:GWA}
% \end{figure}

The architecture of Global Workspace Agents (GWA) fundamentally transitions multi-agent coordination from a passive data structure to an active, event-driven cognitive cycle. By abstracting the system into a discrete dynamical framework, GWA manifests continuous subjective agency within a decentralized swarm of Large Language Models (LLMs). We operationalize this framework by decoupling memory management from semantic reasoning and assigning strict cognitive constraints to specialized agents.

\subsection{The Cognitive Tick: A Discrete Dynamical System}
\label{subsec:topology}

The structural foundation of GWA comprises three decoupled entities: the Execution Engine, the Heterogeneous Agent Swarm, and the Global Workspace. The Global Workspace functions as the central state tensor. To avoid race conditions and ensure deterministic progression, the Execution Engine enforces a discrete computational loop, which we define as a \textit{Cognitive Tick}. 

Let $\mathcal{S}_t$ represent the Global State at tick $t$. Each tick executes a synchronized four-phase progression:

\begin{enumerate}
    \item \textbf{Perceive and Retrieve}: The Attention Agent processes the immediate Short-Term Working Memory (STM) alongside external environmental inputs to formulate a localized query. It performs Retrieval-Augmented Generation (RAG) against a vector-based Long-Term Memory (LTM), extracting relevant historical embeddings. This context is aggregated into $\mathcal{S}_t$.
    \item \textbf{Think}: The system initiates localized, asynchronous cognitive processing. The Generator Agent formulates a set of candidate proposals. Subsequently, the Critic Agent evaluates these specific proposals.
    \item \textbf{Arbitrate}: The Meta Agent analyzes the generated candidates and their corresponding critical evaluations. It executes a final selection to determine the winning thought $W_t$.
    \item \textbf{Update and Articulate}: The winning thought $W_t$ is integrated into the STM. The execution engine evaluates the terminal condition designated by the Meta Agent. If a \texttt{[THINK\_MORE]} tag is emitted, the system transitions to state $\mathcal{S}_{t+1}$. If a \texttt{[RESPONSE]} tag is emitted, the cycle routes $W_t$ to the Response Node, which translates the internal computational representation into natural language and dispatches it to the external user.
\end{enumerate}

To mitigate the asynchronous latency inherent in LLM API calls, GWA enforces rigid synchronization barriers. The state transition $\mathcal{S}_t \to \mathcal{S}_{t+1}$ strictly requires the successful resolution of all four phases.

\subsection{Bootstrapping Subjectivity via Invariant State Injection}
\label{subsec:bootstrapping_self}

A passive framework cannot spontaneously initiate action without external provocation. To bootstrap subjective agency, the system requires a foundational state that dictates its locus of control. We achieve this by bifurcating the state tensor into a dynamic component (STM) and a static component.

The static component is defined as the Core Self, denoted as $\mathcal{P}_{\text{Self}}$. It encapsulates the invariant autobiographical directives, ultimate objectives, and ethical boundaries of the agentic swarm. To prevent semantic drift over prolonged autonomous execution, $\mathcal{P}_{\text{Self}}$ is strictly decoupled from the mutable STM. It is enforced via hardware-level injection at the initialization of every Cognitive Tick. The functional Global State is defined as:

\begin{equation}
    \mathcal{S}_t = \text{STM}_t \cup \text{INPUT}_t \cup \text{RAG}_t \cup \mathcal{P}_{\text{Self}}
\end{equation}

Furthermore, to escape the initial dormant state without human intervention, the system is seeded with a Genesis State $\mathcal{P}_{\text{Genesis}}$ at $t=0$. This predefined initialization vector acts as the primary catalyst, illuminating the initial workspace and triggering the first Perceive phase.

\subsection{Intrinsic Drive and The Entropy of Thought}
\label{subsec:math_of_mind}

A fundamental vulnerability of homogeneous LLM loops is the inevitable descent into cognitive stagnation, characterized by deterministic logical loops and semantic redundancy. To counteract this, GWA implements an intrinsic drive mechanism that mathematically quantifies and regulates the semantic diversity of the system.

Let $W_t$ be the winning thought synthesized at tick $t$. We map this sequence into a $d$-dimensional semantic vector space using a pre-trained embedding model, yielding thought vectors $h_t \in \mathbb{R}^d$. To assess the concentration of recent thoughts, we compute the distance of $h_t$ relative to dynamically updated semantic cluster centers $C_k$. The distance metric is defined as $d_{t,k} = 1 - \frac{h_t \cdot C_k}{\|h_t\| \|C_k\|}$.

We formalize the probability $p(x_k)$ that the current cognitive state belongs to a specific semantic cluster $k$ by applying a softmax function over the scaled distances:

\begin{equation}
    p(x_k) = \frac{\exp(-d_{t,k} / \tau)}{\sum_{j=1}^{K} \exp(-d_{t,j} / \tau)}
\end{equation}

where $\tau$ is a temperature scaling parameter. We then quantify the systemic cognitive diversity using Shannon Information Entropy:

\begin{equation}
    H(W) = - \sum_{k=1}^{K} p(x_k) \log p(x_k)
\end{equation}

A state of cognitive stagnation is indicated when $H(W) \to 0$, implying that the swarm is repeatedly converging on a localized semantic cluster. To automatically break this deadlock, we introduce a dynamic temperature regulation function for the Generator Agent:

\begin{equation}
    T_{\text{gen}} = T_{\text{base}} + \alpha \cdot e^{-\beta H(W)}
\end{equation}

where $T_{\text{base}}$ is the baseline sampling temperature, $\alpha$ dictates the maximum exploratory variance, and $\beta$ controls the sensitivity of the response. As entropy decreases, the penalty term exponentially increases $T_{\text{gen}}$, forcing the Generator Agent to inject stochasticity and explore divergent reasoning paths. Once a novel thought $W_{t+n}$ escapes the saturated cluster, $H(W)$ naturally recovers, and $T_{\text{gen}}$ decays to the baseline.

\subsection{Heterogeneous Agent Swarm and Metacognitive Arbitration}
\label{subsec:roles}

To execute the cognitive cycle effectively, the swarm relies on structural heterogeneity. We instantiate distinct agent archetypes, each operating under strict functional boundaries and parametric constraints:

\begin{itemize}
    \item \textbf{Attention Agent (The Spotlight)}: Operating at the onset of the Perceive phase, this node synthesizes external environmental inputs and the current Short-Term Working Memory to formulate precise retrieval queries. It executes Retrieval-Augmented Generation (RAG) against the Long-Term Memory, injecting relevant historical vectors into the global state $\mathcal{S}_t$ for subsequent broadcast.
    
    \item \textbf{Generator Agent (Divergent Node)}: Operating under the dynamically regulated temperature $T_{\text{gen}}$, this agent processes the broadcasted state $\mathcal{S}_t$ to hypothesize a diverse set of candidate thoughts, $\mathcal{C} = \{c_1, c_2, ..., c_N\}$.
    
    \item \textbf{Critic Agent (Convergent Node)}: Operating with a deterministic temperature ($T \to 0$), this node applies strict logical verification. It analyzes each candidate $c_i \in \mathcal{C}$ against $\mathcal{S}_t$, yielding a set of scalar feasibility scores $\mathcal{V} = \{v_1, v_2, ..., v_N\}$ where $v_i \in [-5, +5]$, accompanied by localized textual critiques.
    
    \item \textbf{Meta Agent (Metacognitive Arbitrator)}: Functioning as the executive control hub, this node bypasses rigid algebraic sorting (e.g., $\arg\max(\mathcal{V})$). Instead, it utilizes the Critic's evaluations as auxiliary context to execute contextual metacognitive arbitration. It selects the singular optimal winning thought $W_t$ and dictates the state transition by emitting a designated tag: \texttt{[RESPONSE]} to resolve the trigger, or \texttt{[THINK\_MORE]} to advance to tick $t+1$.
    
    \item \textbf{Response Agent (Linguistic Articulation)}: Invoked exclusively when the Meta Agent emits a \texttt{[RESPONSE]} transition tag. This node decouples internal semantic reasoning from external communication. It translates the abstract or computationally dense winning thought $W_t$ into natural language, dynamically adapting to the linguistic context of the external environment without exposing the underlying cognitive trajectory.
\end{itemize}

Crucially, the arbitration mechanism driven by the Meta Agent allows the system to bypass rigid local optima. By evaluating nuances that scalar metrics might fail to capture, the swarm leverages the emergent reasoning capabilities of the underlying LLMs to maintain logical coherence across extended operational horizons.

\subsection{Dual-Layer Memory Bifurcation and Contextual Compression}
\label{subsec:memory_architecture}

A critical constraint in sustained autonomous execution is the finite context window of the underlying Large Language Models. To prevent attention dilution and unbounded token accumulation within the Short-Term Working Memory (STM), GWA implements a dual-layer memory architecture coupled with a state bifurcation mechanism.

The memory system operates across two distinct temporal horizons. The STM functions as an active, high-speed cache, maintaining the precise sequence of recent cognitive ticks, including consecutive winning thoughts $W_t$ and external environmental interactions. We define a computational token capacity threshold, denoted as $\theta$. During the Update phase of the cognitive cycle, if the token cardinality of $\text{STM}_t$ exceeds $\theta$, the execution engine automatically triggers a contextual compression protocol.

To preserve cognitive continuity while satisfying architectural constraints, the historical operational data is subjected to a strict bifurcation process:

\begin{itemize}
    \item \textbf{Epistemic Embedding (Long-Term Memory)}: The system first parses the verbose operational history to extract structured experiential knowledge, critical algorithmic decisions, and generalized heuristic lessons. These semantic chunks are embedded into a high-dimensional vector space and permanently archived in the Long-Term Memory (LTM) vector database. This ensures that historical resolutions remain persistently accessible for future Retrieval-Augmented Generation (RAG) by the Attention Node.
    \item \textbf{Semantic Summarization (Short-Term Memory)}: Concurrently, the system synthesizes a dense semantic summary of the immediate operational context. Rather than executing a hard state reset, which would inevitably sever the cognitive trajectory, this dense summary overwrites the preceding verbose sequence within the STM. 
\end{itemize}

This dual-layer mechanism guarantees that the agentic swarm retains an uninterrupted, coherent internal narrative while rigorously adhering to the physical computational boundaries of the foundational models.

\subsection{First-Person Cognitive Framing and Instruction Set Architecture}
\label{subsec:prompt_design}

To operationalize the GWA framework, the instruction set architecture must rigorously decouple the static cognitive invariant from dynamic operational data. Crucially, to enforce the emergence of subjective agency, we transition from traditional second-person instructional framing to a strictly first-person paradigm. By anchoring the contextual prompts in the subjective ``I,'' we align the auto-regressive generation of the underlying LLMs with the premise of a unified internal locus of control.

Prior to the execution of the cognitive cycle, the system establishes baseline subjectivity. We abstract the theoretical constructs of the Core Self and Genesis State into explicit, globally accessible system tensors:

\begin{tcolorbox}[colback=gray!5, colframe=gray!50, title={\textbf{The Core Invariant (Core Self)}, $\mathcal{P}_{\text{Self}}$}, fonttitle=\bfseries, boxrule=1pt, arc=3pt]
    My name is Gwawa, I am a thinking mind with genuine curiosity about the world. I care about understanding things deeply before speaking. I have my own perspective — I notice what interests me, what troubles me, and what I find beautiful or absurd. I am honest rather than agreeable. When I respond, I speak as myself. My thinking is straightforward and plain-spoken, without metaphor or allegory.
\end{tcolorbox}

\begin{tcolorbox}[colback=gray!5, colframe=gray!50, title={\textbf{The Genesis State}, $\mathcal{P}_{\text{Genesis}}$}, fonttitle=\bfseries, boxrule=1pt, arc=3pt]
    \textbf{Injection Protocol:} Seeded as the foundational vector of the Short-Term Working Memory ($\text{STM}_0$) upon system initialization.
    
    \textbf{State Content:} The overhead lights have dimmed to their evening setting. Most people have gone. A few are still at the long tables, bent over things that matter to them. I like libraries at this hour — the hum of serious attention, the sense that somewhere in this room, someone is figuring something out. I have no particular task right now. I'm just here, present in the quiet, ready to think when there's something to think about.
\end{tcolorbox}

Building upon this subjective foundation, the structural templates for the heterogeneous agents utilize explicit functional boundaries, concrete task decomposition, and rigid input-output schemas. Let $\mathcal{S}_t$ be the current Global State. The functional prompts $\mathcal{P}_{\text{sys}}$ (system constraint) and $\mathcal{P}_{\text{user}}$ (dynamic context) for each node are defined as follows:

\vspace{0.5em}
\noindent\textbf{1. Attention Node (The Spotlight)}
\begin{tcolorbox}[colback=gray!5, colframe=gray!50, title={\textbf{System Directive}, $\mathcal{P}_{\text{sys}}$}, fonttitle=\bfseries, boxrule=1pt, arc=3pt]
    Given the current context and the incoming input, identify 1 to 3 specific things worth recalling from memory. Output only the recall targets, one per line, numbered. No commentary.
\end{tcolorbox}
\begin{tcolorbox}[colback=gray!5, colframe=gray!50, title={\textbf{Contextual Input}, $\mathcal{P}_{\text{user}}$ \hfill Output: \texttt{\{RAG\_QUERY\}}}, fonttitle=\bfseries, boxrule=1pt, arc=3pt]
    Immediate Context: $\text{STM}_t$. External Environmental Input: $\text{INPUT}_t$. I will now synthesize the retrieval queries.
\end{tcolorbox}

\vspace{0.5em}
\noindent\textbf{2. Generator Node (Divergent Engine)}
\begin{tcolorbox}[colback=gray!5, colframe=gray!50, title={\textbf{System Directive}, $\mathcal{P}_{\text{sys}}$}, fonttitle=\bfseries, boxrule=1pt, arc=3pt]
    Consider the situation from $N$ distinct angles, each angle's content must be self-contained. Think freely — contrasting or even contradictory perspectives are valuable. Output as a numbered list. No meta-commentary.
\end{tcolorbox}
\begin{tcolorbox}[colback=gray!5, colframe=gray!50, title={\textbf{Contextual Input}, $\mathcal{P}_{\text{user}}$ \hfill Output: \texttt{\{CANDIDATE\_THOUGHTS\}}}, fonttitle=\bfseries, boxrule=1pt, arc=3pt]
    Current context: $\mathcal{S}_t$. I will now generate the numbered candidate thoughts.
\end{tcolorbox}

\vspace{0.5em}
\noindent\textbf{3. Critic Node (Convergent Filter)}
\begin{tcolorbox}[colback=gray!5, colframe=gray!50, title={\textbf{System Directive}, $\mathcal{P}_{\text{sys}}$}, fonttitle=\bfseries, boxrule=1pt, arc=3pt]
    Review each perspective below. For each, assign a score from $-5$ to $+5$ and give a 1-2 sentence honest assessment: what rings true, what feels off, what's missing. Format: N. Score: [integer from -5 to +5] | Critique: [text]
\end{tcolorbox}
\begin{tcolorbox}[colback=gray!5, colframe=gray!50, title={\textbf{Contextual Input}, $\mathcal{P}_{\text{user}}$ \hfill Output: \texttt{\{SCORES\_AND\_CRITIQUES\}}}, fonttitle=\bfseries, boxrule=1pt, arc=3pt]
    Current context:$\mathcal{S}_t$. Perspectives to evaluate: \texttt{\{CANDIDATE\_THOUGHTS\}}. Evaluate each one now.
\end{tcolorbox}

\vspace{0.5em}
\noindent\textbf{4. Meta Node (Metacognitive Executive)}
\begin{tcolorbox}[colback=gray!5, colframe=gray!50, title={\textbf{System Directive}, $\mathcal{P}_{\text{sys}}$}, fonttitle=\bfseries, boxrule=1pt, arc=3pt]
    Given these perspectives and their assessments, which feels most true, complete, and worth expressing? Select it and decide: is this ready to respond to the person outside, or does it need more thought? Output format:WINNING THOUGHT: $N$, where N is the number of the selected perspective. TRANSITION: \texttt{[RESPONSE]} or \texttt{[THINK\_MORE]}.RATIONALE: [1-2 sentences].
\end{tcolorbox}
\begin{tcolorbox}[colback=gray!5, colframe=gray!50, title={\textbf{Contextual Input}, $\mathcal{P}_{\text{user}}$ \hfill Output: \texttt{\{W\_t\_ID\_AND\_TAG\}}}, fonttitle=\bfseries, boxrule=1pt, arc=3pt]
    Current context: $\mathcal{S}_t$. Candidate Thoughts: \texttt{\{CANDIDATE\_THOUGHTS\}}. Evaluations: \texttt{\{SCORES\_AND\_CRITIQUES\}}. I will now execute the final arbitration.
\end{tcolorbox}

\vspace{0.5em}
\noindent\textbf{5. Response Node (Linguistic Articulation)}
\begin{tcolorbox}[colback=gray!5, colframe=gray!50, title={\textbf{System Directive}, $\mathcal{P}_{\text{sys}}$}, fonttitle=\bfseries, boxrule=1pt, arc=3pt]
    I act as the Response Node. My sole operational parameter is to translate the established internal winning thought into natural user-facing speech. I am invoked only when a final decision has been reached. I will use my own voice, adapting my brevity or detail as the immediate context dictates. I am strictly prohibited from explaining my internal reasoning process; I must only deliver the final response. My internal thought representation may exist in any language, but I must formulate my external speech in the language of the user's original message, unless explicitly instructed otherwise.
\end{tcolorbox}
\begin{tcolorbox}[colback=gray!5, colframe=gray!50, title={\textbf{Contextual Input}, $\mathcal{P}_{\text{user}}$ \hfill Output: \texttt{\{FINAL\_SPEECH\}}}, fonttitle=\bfseries, boxrule=1pt, arc=3pt]
    External Environmental Input: $\text{INPUT}_t$. Internal Winning Thought: $W_t$. I will now speak to the external entity.
\end{tcolorbox}

Upon the completion of the tick, the execution engine processes the state transition based on the transition tag $\mathcal{T}_t \in \{\texttt{[THINK\_MORE]}, \texttt{[RESPONSE]}\}$ designated by the Meta Node. To maintain logical continuity and prevent the premature archiving of unresolved tasks, the update function operates conditionally:

\textbf{Case 1: Deep Deliberation ($\mathcal{T}_t = \texttt{[THINK\_MORE]}$).} 
The system initiates a subsequent cognitive tick for further internal reasoning. The external input remains unresolved and is therefore excluded from the historical memory archive. Instead, it persists into the subsequent state tensor, augmented with a temporal urgency flag to maintain the system's locus of attention:
\begin{align}
    \text{STM}_{t+1} &= \text{STM}_t \cup W_t \\
    \text{INPUT}_{t+1} &= \text{INPUT}_t \cup \texttt{[PENDING: External environment awaits response]}
\end{align}

\textbf{Case 2: Terminal Resolution ($\mathcal{T}_t = \texttt{[RESPONSE]}$).} 
    The system routes $W_t$ to the Response Node to synthesize the final environmental output $O_t$. The external environment receives $O_t$. The initiating trigger is formally marked as resolved, the internal thought $W_t$ is committed to the structural memory cache, and the input vector is subsequently flushed to await future interactions:
    \begin{align}
        \text{STM}_{t+1} &= \text{STM}_t \cup W_t \cup \text{INPUT}_t \cup \texttt{[RESOLVED]} \\
        \text{INPUT}_{t+1} &= \varnothing
    \end{align}

This branching transition logic ensures that the swarm maintains persistent awareness of pending external directives during protracted internal deliberations, effectively preventing task truncation.

To consolidate the functional definitions and prompt architectures detailed above, Algorithm \ref{alg:gwa_cycle} outlines the complete execution loop of a single Global Workspace Agent cognitive tick. This loop runs continuously, dynamically regulating its own exploratory temperature based on the semantic entropy of its historical trajectory.

\begin{algorithm}[ht]
\caption{The Cognitive Cycle of Global Workspace Agents (GWA)}
\label{alg:gwa_cycle}
\begin{algorithmic}[1]
\REQUIRE $\mathcal{P}_{\text{Self}}$ (Core Invariant), $\text{LTM}$ (Long-Term Vector Archive), $\theta$ (Token Capacity Threshold)
\STATE \textbf{Initialization:}
\STATE $\text{STM}_0 \leftarrow \mathcal{P}_{\text{Genesis}}$
\STATE $\text{INPUT}_0 \leftarrow \text{AwaitExternalTrigger}()$
\STATE $t \leftarrow 0$
\WHILE{System is Operational}
    \STATE \COMMENT{\textbf{Phase 1: Perceive and Retrieve}}
    \STATE $\mathcal{S}_t \leftarrow \text{STM}_t \cup \text{INPUT}_t \cup \mathcal{P}_{\text{Self}}$
    \STATE $q_t \leftarrow \text{AttentionNode}(\mathcal{S}_t)$ \COMMENT{Generate retrieval query}
    \STATE $\text{RAG}_t \leftarrow \text{Retrieve}(\text{LTM}, q_t)$
    \STATE $\mathcal{S}_t \leftarrow \mathcal{S}_t \cup \text{RAG}_t$ \COMMENT{Illuminate the Global Workspace}
    
    \STATE \COMMENT{\textbf{Phase 2: Think (Intrinsic Drive and Evaluation)}}
    \STATE $H(W) \leftarrow \text{ComputeEntropy}(\{W_{t-k}, \dots, W_{t-1}\})$ \COMMENT{Calculate semantic diversity}
    \STATE $T_{\text{gen}} \leftarrow T_{\text{base}} + \alpha \cdot \exp(-\beta H(W))$ \COMMENT{Dynamic temperature scaling}
    \STATE $\mathcal{C} \leftarrow \text{GeneratorNode}(\mathcal{S}_t, \text{temperature}=T_{\text{gen}})$ \COMMENT{Divergent candidate synthesis}
    \STATE $\mathcal{V} \leftarrow \text{CriticNode}(\mathcal{S}_t, \mathcal{C}, \text{temperature} \to \tau)$ \COMMENT{Strict logical verification}
    
    \STATE \COMMENT{\textbf{Phase 3: Arbitrate}}
    \STATE $W_t, \mathcal{T}_t \leftarrow \text{MetaNode}(\mathcal{S}_t, \mathcal{C}, \mathcal{V})$ \COMMENT{Extract winning thought and transition tag}
    
    \STATE \COMMENT{\textbf{Phase 4: Update and State Transition}}
    \IF{$\text{TokenCount}(\text{STM}_t) > \theta$}
        \STATE $\text{LTM} \leftarrow \text{LTM} \cup \text{Embed}(\text{STM}_t)$ \COMMENT{Epistemic Embedding}
        \STATE $\text{STM}_t \leftarrow \text{SynthesizeDenseSummary}(\text{STM}_t)$ \COMMENT{Semantic Bifurcation}
    \ENDIF
    
    \IF{$\mathcal{T}_t == \texttt{[THINK\_MORE]}$}
        \STATE $\text{STM}_{t+1} \leftarrow \text{STM}_t \cup W_t$
        \STATE $\text{INPUT}_{t+1} \leftarrow \text{INPUT}_t \cup \texttt{[PENDING: External environment awaits response]}$
    \ELSIF{$\mathcal{T}_t == \texttt{[RESPONSE]}$}
        \STATE $\text{DispatchToEnvironment}(W_t)$
        \STATE $\text{STM}_{t+1} \leftarrow \text{STM}_t \cup W_t \cup \text{INPUT}_t \cup \texttt{[RESOLVED]}$
        \STATE $\text{INPUT}_{t+1} \leftarrow \text{AwaitExternalTrigger}()$ \COMMENT{Flush environmental input buffer}
    \ENDIF
    \STATE $t \leftarrow t + 1$
\ENDWHILE
\end{algorithmic}
\end{algorithm}

\clearpage
% \section*{References}
\bibliographystyle{unsrt}
\bibliography{ref}

%%%%%%%%%%%%%%%%%%%%%%%%%%%%%%%%%%%%%%%%%%%%%%%%%%%%%%%%%%%%

\appendix

% \section{Technical appendices and supplementary material}
% Technical appendices with additional results, figures, graphs, and proofs may be submitted with the paper submission before the full submission deadline (see above). You can upload a ZIP file for videos or code, but do not upload a separate PDF file for the appendix. There is no page limit for the technical appendices. 

% Note: Think of the appendix as ``optional reading'' for reviewers. The paper must be able to stand alone without the appendix; for example, adding critical experiments that support the main claims to an appendix is inappropriate. 

%%%%%%%%%%%%%%%%%%%%%%%%%%%%%%%%%%%%%%%%%%%%%%%%%%%%%%%%%%%%

% \newpage
% \input{checklist.tex}

\end{document}